\documentclass[10pt]{article}

\usepackage{graphics}
\usepackage{graphicx}

\usepackage[a4paper, left=35mm,right=35mm,top=34mm,bottom=34mm]{geometry}
\usepackage[utf8]{inputenc}
\usepackage[T1]{fontenc}
\usepackage[english]{babel}

\usepackage{enumerate}
\usepackage{graphicx}
\usepackage{hyperref}
\hypersetup{
    colorlinks=true,
    linkcolor=blue,
    filecolor=magenta,      
    urlcolor=cyan,
}
\usepackage{listings}
\usepackage{color}

\definecolor{dkgreen}{rgb}{0,0.6,0}
\definecolor{gray}{rgb}{0.5,0.5,0.5}
\definecolor{mauve}{rgb}{0.58,0,0.82}
\lstdefinelanguage{MRGC++}{%
  language=C++,
  morekeywords={T, U, MPI_Irecv, MPI_Isend, MPI_Allreduce, MPI_Waitall, Compute, Map, abs, max, Swap, MPI_Recv_init, MPI_Send_init, MPI_Startall, Copy, Init, InitRecv, InitSend, InitAllReduce, Send, Recv, AllReduce, Finalize, InitSnapshot, Snapshot, SwitchAsync, SnapReduce, MPI_Test, MPI_Start}
}
\usepackage{mathtools,amsthm,amssymb,amsfonts}
\usepackage{algorithm}
\usepackage{algorithmic}
\makeatother
\theoremstyle{plain}

\theoremstyle{definition}

\theoremstyle{remark}

\usepackage{caption} 
\usepackage{fancyhdr}

\author{
  {\normalsize Guillaume Gbikpi-Benissan}\thanks{Ecole Centrale Paris, France
    (correspondence, frederic.magoules@hotmail.com).}
  \and
  {\normalsize R\'emi Cerise}\footnotemark[1]
  \and
  {\normalsize Patrick Callet}\footnotemark[1]
  \and
  {\normalsize Fr\'ed\'eric Magoul\`es}\footnotemark[1]
}
\title{Spectral Domain Decomposition Method for Natural Lighting and Medieval Glass Rendering}
\date{}

\begin{document}
\maketitle
\thispagestyle{fancy}

\begin{abstract}
\noindent In this paper, we use an original ray-tracing domain decomposition method to address image rendering of naturally lighted scenes.
This new method allows to particularly analyze rendering problems on parallel architectures, in the case of interactions between light-rays and glass material.
Numerical experiments, for medieval glass rendering within the church of the Royaumont abbey, illustrate the performance of the proposed ray-tracing domain decomposition method (DDM) on multi-cores and multi-processors architectures. On one hand, applying domain decomposition techniques increases speedups obtained by parallelizing the computation. On the other hand, for a fixed number of parallel processes, we notice that speedups increase as the number of sub-domains do.
\end{abstract}


\section{Introduction}

This work is about a study of ancient stained glass windows as they might be used in the Cistercian art in France in the middle of the XIIIth century. Using a model of the abbey church of Royaumont, optical simulations are done with Virtuelium, a physically-based rendering open source software developed at Ecole Centrale Paris (France). More specifically, our interest is to produce consistent visualization based on some hypotheses about the visual appearance of Cistercian stained glass windows as they could have been seen in the church around 1250 AD. We use several samples of medieval glasses for this study, many of them coming from the abbey of Maubuisson, which is very close to Royaumont in terms of geographic location and time. As well, the methodology required more recent glasses, although manufactured in a very traditional way, like the complete set of Saint-Just glasses (Saint-Gobain group) we also used.

Interactions between natural lighting and medieval glass materials are hard to render in their whole complexity because of their heterogeneity in terms of optical behavior. For instance, measurements on glass tiles revealed a non uniform thickness and the presence of micro particles of air inside the tiles. That is responsible of an irregular volume scattering of light. Traditional physically-based rendering engines~\cite{Pharr:2010:PBR:1854996, LuxRender, Shirley:2012:BPR:2407783.2407785} use extrinsic properties of materials, such as spectral reflectances and transmittances. These approaches are not complete enough to efficiently address our problem. An originality of the Virtuelium approach consists in coupling both extrinsic and intrinsic properties (optical constants).

The specific methodology of acquisition on glass samples is thus described in the first section. Section 2 presents some rendering algorithms including those used in Virtuelium, then we give details about how these algorithms are computed in parallel. Finally, a presentation of resulted images and speedups will close our demonstration.

\section{Data acquisition}

Extrinsic and intrinsic properties of glass materials can be measured with two distinct methods. Spectrophotometry~\cite{bass1995handbook} is commonly used to acquire spectral responses in reflection or transmission. This approach mainly consists in lighting the surface to be measured with known spectral (the emission spectrum) and geometrical (the incident angle) conditions. By analyzing the energetic quantities measured along a wavelength range, we can deduce the spectral functions for reflectance or transmittance. Then, the material is fully characterized in one point of its surface by repeated single measurements for regular sampling of both incident and view angles. In the case of glass materials, the obtained function is called Bidirectional Transmittance Distribution Function (BTDF) and it belongs to a larger family of spectral distribution functions.

Another method based on spectroscopic ellipsometry can be used to determine a BTDF. This second method offers a way to measure the complex index of refraction
\begin{equation}
\label{eq:optical}
\tilde{n}(\lambda) = n(\lambda)+ i k(\lambda) = n(\lambda) (1 + i \kappa(\lambda))
\end{equation}
which is called  "optical constants"~\cite{palik1985handbook, callet1998couleur}.
The quantities $n(\lambda)$ denotes the optical index, $k(\lambda)$ stands for the index of absorption and $\lambda$ represents the wavelength. At the opposite of extrinsic properties, optical constants really define the electronic behavior of dielectric materials, and not just a spectral response. Besides, they are often used to simulate the visual appearance of metallic surfaces~\cite{cgaBerger, Woollam199444, 5111815} as these materials fully satisfy the Fresnel conditions (non-scattering and homogeneous). But that is not true for glass materials and particularly medieval glass materials.

In order to deal with the heterogeneity of medieval glass materials, input data are measured from a piece of a height-hundred-years-old stained glass window coming from the Maubuisson abbey. On another hand, a library of modern samples (from Saint Just Corporation) with correct Fresnel conditions are also used. Then, by respecting a rigorous colorimetric protocol, we select appropriate modern glass materials which are visually close to the ancient one. More details can be found in~\cite{Cerise:2012:NLM:2426256.2426337}.

\section{Rendering algorithms}

To simplify next discussions, we will only consider equations for the Bidirectional Reflectance Distribution Function (BRDF). The reasoning for the BTDF is the same except that we do not only consider a dome for incident lights but the entire sphere since the studied surface is non-opaque. The global rendering equation is deduced from the Radiative Transfer Equation (RTE) and described by Kajiya~\cite{Kajiya:1986:RE:15886.15902} as follow:
\begin{equation}
\label{eq:radiative_theo}
L_{r}(\vec{\omega_{o}}) = \int_{\Omega} F_{r}(\vec{\omega_{i}}, \vec{\omega_{o}}) L_{i}(\vec{\omega_{i}}) \vec{n}.\vec{w_{i}} d \omega_{i}
\end{equation}
where $\Omega$ is the dome of incident lights. Literally, a radiative balance is calculated. The remitted light $L_{r}$ in a direction $\vec{\omega_{o}}$ depends on every incident light $L_{i}$ emitted by the dome. It is then sufficient to know the BRDF, $F_{r}$, for computing the remitted light from a given incident light. In an applicative way, the equation (\ref{eq:radiative_theo}) can be simplified as follow if we consider only point or directional light-sources:
\begin{equation}
\label{eq:radiative_appli}
L_{r}(\vec{\omega_{o}}) = \sum_{s 1}^N F_{r}(\vec{\omega_{s}}, \vec{\omega_{o}}) L_{s}(\vec{\omega_{s}}) \vec{n}.\vec{w_{s}}
\end{equation}
where $N$ is the number of light-sources, $L_{s}$, the emission spectrum of the current light-source $s$ and $\vec{\omega_{s}}$ the incident direction.

At the scope of a whole 3D scene, image rendering engines observe two major kinds of interaction between lights and geometric objects. We talk about local illumination when the only lights hitting an object are those directly coming from the light sources without any other interaction. At the opposite, the mutual contributions of all the objects present in the 3D scene is then responsible of global illumination. As a result, rendering algorithms can be sorted between those which compute global illumination and those which ignore it. For now, Virtuelium provides one algorithm of each kind: the "Scanline rendering" technique for local illumination and the "Photon Mapping" for global illumination.

\subsection{Without global illumination}

The "Scanline rendering"~\cite{Wylie:1967:HPD:1465611.1465619} is based on the idea of inverse ray tracing algorithm~\cite{Arvo86backwardray}. Indeed it is easier not to traverse light paths in the logic direction, but from the camera to light-sources. As the image to be computed can be viewed as a matrix of pixels, a light-ray is emitted from each pixel, orthogonally to the image plane. Once emitted, each light-ray evolves independently. When a ray intersects with the closest object on his road, two actions can be executed. In a first time, we have to evaluate the received luminance at the given viewed direction. In order to achieve this goal, new rays are shot from the hit point to each light-source, thus determining all the needed incident directions. Then, secondary rays are thrown regarding to reflection and/or refraction laws and the process is repeated. The algorithm stops when the energetic value attached to the ray goes bellow a threshold or after the ray has bounced a predetermined number of times. Additional information retrieved from a hit point also allows more complex computations. By example, this can be texture coordinates for spatial distribution maps.

The complexity of ray tracers are mostly contained in the complexity of intersectors. That is why the main difference between algorithms of this family resides in the way polygons of objects are sorted. Sort in "Scanline rendering" is achieved by projecting every polygons onto the image plane. Then, the image is computed line by line, from top to bottom, determining the color of each pixel by considering the closest polygons around. Other very common algorithm also exists (see the "Z-buffer technique~\cite{Catmull:1974:SAC:907242} which is nowadays implemented by default on graphic cards). The main advantage of the "Scanline rendering" is that each pixel is only evaluated once. In return, the memory cost is high because all the polygons of the scene must be loaded at the same time, leading to bad performances for scenes with complex geometries.

\subsection{With global illumination}

Including global illumination (GI) in image rendering processes is a major progress in the quest of photo-realism, but great differences exist between GI techniques. For example, "Radiosity" methods~\cite{Wallace:1987:TSR:37402.37438, Sillion:1989:GTM:74333.74368} transform the phenomenon of global illumination into a system of linear equations, with different ways of resolutions: direct resolution~\cite{journals:vc:BuD89} (very effective but with a high complexity) or iterative algorithms~\cite{Cohen:1988:PRA:54852.378487}. In another direction, the use of the stochastic algorithm of "Monte-Carlo"~\cite{134595} is sometimes used despite its slower convergence. "Path Tracing" methods~\cite{CGF:CGF1863} launch random rays from pixels of the image plane until one hits an object. It can be bi-directional (shots rays from camera and sources simultaneously). "Metropolis Light Transport" algorithm (MTL)~\cite{Hachisuka:2008:PPM:1409060.1409083} optimizes "Path Tracing" by replacing the random shooting with heuristics. Modern versions of theses algorithms are progressive. 
This means they are not limited by a maximum number of bounces but instead continue to converge while the user allows it (or until a quality condition is reached). 

The version of "Photon Mapping" algorithm implemented in Virtuelium is not progressive yet but is simpler to implement. It was first defined by Jensen in 1996 and is improved since this date~\cite{Jensen:1996:GIU:275458.275461, Jensen:2004:PGG:1103900.1103920}.
It decomposes the rendering process into two steps which are executed sequentially. In the pre-rendering step, the position of photons (light-rays launched from light-source) hitting objects are stored in several appropriate structures (photon maps). At least two photon maps are needed, one for the global illumination itself and one for caustics. During the next step, four different contributions are evaluated based on the fact that $L_{r}(\vec{\omega_{o}})$ can be decomposed into a sum of different integrals. First, the direct and specular contributions are computed the same way than in "Scanline Rendering". Then, the caustic and indirect diffuse contributions are deduced from the two photon maps. Most recent versions of Photon Mapping are nevertheless progressive~\cite{Hachisuka:2008:PPM:1409060.1409083, Hachisuka:2011:RAP:2019627.2019633}.

\section{Standard parallelization}

The most common way of parallelizing an image rendering algorithm consists in decomposing the image grid. Indeed, as each pixel can always be treated separately without any interaction, it is simple to distribute the computations of several of them over multiple CPUs for a full asynchronous execution. However, some area of the image are probably longer to be rendered than others because of the heterogeneous spacing of objects, materials and lights-sources in the scene. 
Thus, the first optimization to bring is a dynamic job-balancing mechanism allowing faster threads to work more and ensuring there is no inactivity period for any of them. Something similar can be applied to the global illumination pre-rendering step, replacing the image grid by the list of light-sources. Besides, because many photons are shot in several directions from the sources, rays can also be distributed dynamically in order to maintain a constant activity on every threads.

The main problem with this fully distributed solution is linked to the fact that the whole scene geometry must be known by each computational node. Predetermining the whole light path of a ray is obviously nearly impossible. Furthermore, each polygons in the scene can be hit several times by different rays. For these reasons, every polygon has to be copied onto the memory of each thread, what is not optimal. Hybrid computing (distributed + shared memory) can be used to assure that only a single copy exists on a computational node (and not one per thread), but the problem remains on multiple-node architectures. We know that copying modern complex scenes with several millions of polygons can become a limitation factor, depending on the network bandwidth.

When dealing with full spectral rendering engines like Virtuelium, another idea which is simple to implement consists in decomposing the spectral data themselves. Whereas RGB or RGBA values have only 3 or 4 components, size spectral values used in Virtuelium are often fixed to an array of 81 scalar values encapsulating the visible part of the wavelength range (380 to 780 nm with sampling step of 5 nm). But previous rendering techniques always consider wavelength separately. That is why it is possible to cut the spectral array by group of $n$ sub-values (for instance, $n$ can be equal to 3 of 4 to return to something very close to the RGB world). Only fluorescent phenomena bring interactions between wavelengths but this problem is solved if we gather wavelengths craftily to take into account this physical law. Nevertheless, the same problems remains and the scene geometry always has to be copied on every computational node.

The solution we proposed is thus to apply the ray-tracing domain decomposition method introduces in~\cite{magoules:patent:2011}.

\section{Domain Decomposition Method}

By splitting a global domain into several small sub-domains, domain decomposition methods~\cite{SBG1996}, \cite{QV1999}, \cite{TW2005}, \cite{Jar2007} allow to load input data and gather results in a parallel way, as each sub-domain can be associated to a unique processor. This splitting can be done only once for multiple execution of the processing algorithm. That is an important advantage particularly when considering very large models besides the fact that in such case, problems of memory allocation are avoided if the sub-domains are small enough.

A basic approach consists in simply splitting the set of pixels into multiple sets, one per sub-domain. In each sub-domain the light rays are shot in the whole geometry. A great part of the work is duplicated, especially the intersections detection and the loading/preprocessing of the model. Since there is absolutely no communication between the processes, this is a good candidate for largely distributed systems.

However, this method raises some load balancing issues since the processing time of each sub-domain can vary a lot. Thus there is another idea consisting in loading sub-domains on demand but it requires to compute a hierarchical acceleration structure in order to only load the first levels at the start. During the traversal of this hierarchical structure the data corresponding to a node are loaded when the node is reached. If it is an interior node, these data are the child nodes information, if it is a leaf, the data is the corresponding mesh and materials data. This system is pretty complicated to implement and needs a specific pre-computed data structure and it would be necessary to cover the latency of the node loading. Such a method is useful in a context where light rays launched by a process don't spread out of a specific bounded part of the model. Otherwise the process would tend to load the complete geometry.

The method described in~\cite{magoules:patent:2011} takes more advantage of some efficient domain decomposition techniques~\cite{magoules:journal-auth:4}, \cite{magoules:journal-auth:16}, \cite{magoules:journal-auth:21}. Besides the splitting of the global geometry itself, information along interfaces is shared between computational units which are processing neighboring sub-domains. A continuous approach~\cite{Des1993}, \cite{Gha1997}, \cite{CN1998}, \cite{magoules:journal-auth:28}, \cite{magoules:journal-auth:23}, \cite{magoules:journal-auth:18}, \cite{magoules:journal-auth:14} can be used to design efficient interface conditions.
Similarly a discrete approach~\cite{magoules:journal-auth:8}, \cite{magoules:proceedings-auth:6}, \cite{magoules:journal-auth:29}, \cite{magoules:journal-auth:12}, \cite{magoules:journal-auth:20} 
can be used, which may increase significantly the performance of the algorithm.
The link between the continuous and discrete interface condition can be established like in~\cite{magoules:journal-auth:17}.

In this work, we split the geometry of the model into multiple sub-domains and base our method on the domain decomposition method~\cite{magoules:journal-auth:24}, \cite{magoules:journal-auth:9}, \cite{magoules:journal-auth:10} where the interface conditions assure the continuity of the light ray properties (such as direction, amplitude, angle, etc.) from one sub-domain to another one, as detailed in~\cite{magoules:patent:2011}. We only analyze the rays passing through interfaces between sub-domains. From a processor point of view, one could replace all neighboring sub-domains model by a simplified version in order to be more efficient.

Yet, unlike classical domain decomposition methods, a computational unit does not process only one sub-domain. Our concern here is related to the load balancing which could be very bad, for instance if there were only one light-source. In such case, most of the workload would be held by the unit processing the sub-domain containing the light-source. This is why we use a less static load-balancing scheme. From a processor point of view, the idea is to start by loading a certain number of sub-domains according to memory limitation. When it remains few light rays to be handled in a sub-domain, this sub-domain is unloaded if there is still other currently not handled sub-domains with a lot of rays not processed. Then the processor starts loading one or more of these sub-domains while handling another sub-domain already available in the memory. Unloading sub-domains allows doing most of the results gathering during the processing of other rays. This overlapping gathering and processing is efficient since gathering mainly uses the communication system. One could find a more complete description of an efficient implementation in~\cite{magoules:patent:2011}.

\section{Results and discussions}

\begin{figure}
\centering
\scalebox{1.0}{\includegraphics[width=0.45\textwidth]{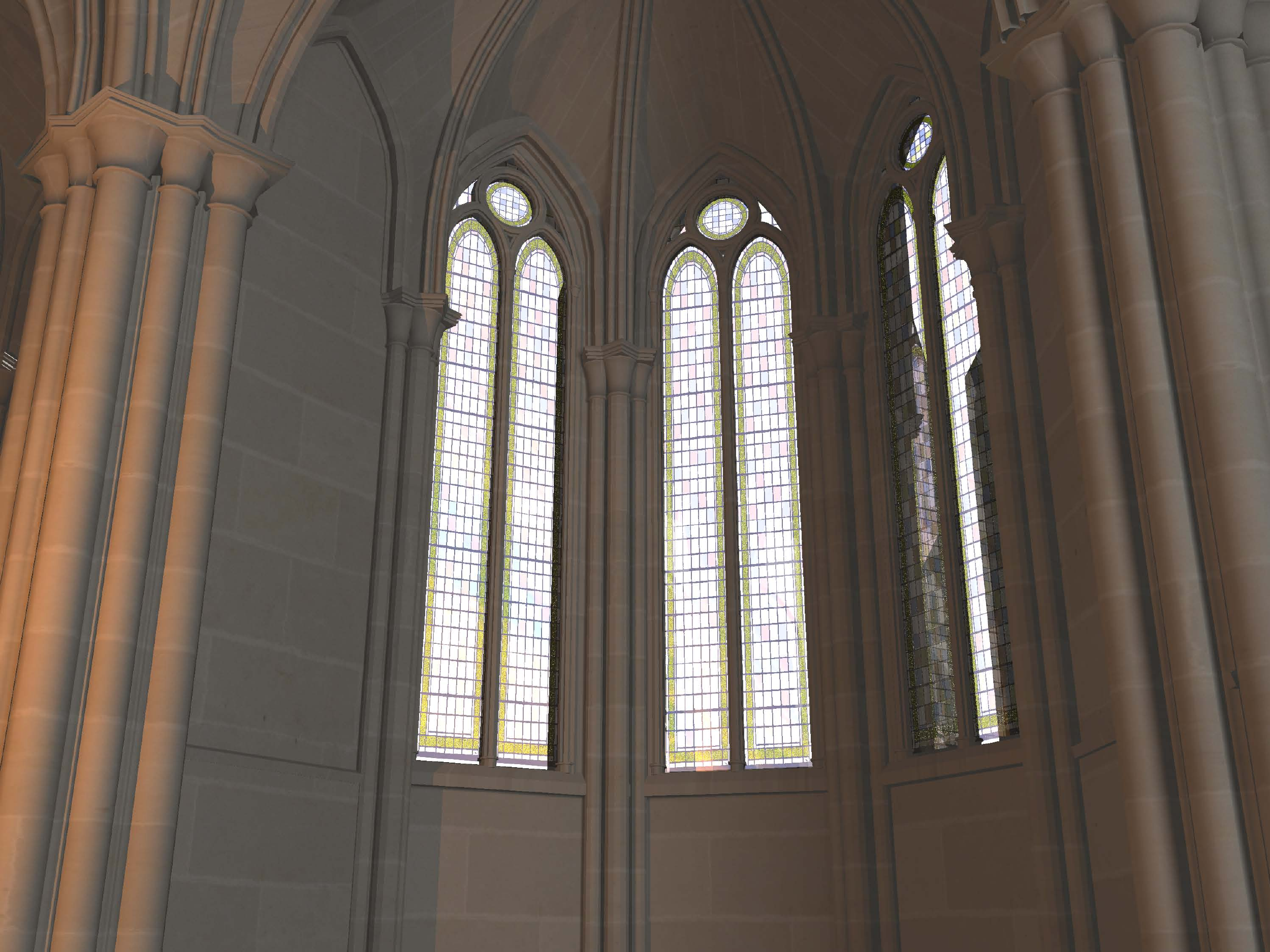}}
\caption{Illustration of the image rendering in the church of Royaumont abbey (interior view)}
\label{fig:virtuelium01}
\end{figure}

An image rendered with Virtuelium is presented in figure~\ref{fig:virtuelium01}. Although it is not fully viewable, the whole church model has been rendered using our new parallelization method. The model we used for representing stained glass windows is quite particular. Indeed, instead of directly applying a texture, we used a distribution map of optical constants. According to medieval Cistercian art, nearly clear glass materials were used rather than highly colored tiles. Nevertheless, with our distribution map, we can create diversity between tiles, what is particularly visible on the right windows (where reflections of architecture and other windows are visible). The yellow edgings are also created only with the map. Distribution maps can also be used to distribute other information along the object geometry. This is the case here for glass thicknesses. Only refractive properties of glass materials are simulated for now. One of our future objectives is to develop the material model in order to extend our simulations to a larger part of the physical phenomenon (by example, light scattering is needed). Again, new distribution maps could be used to accurately represent the material complexity.

\begin{table}
\centering
{\small
\begin{tabular}{|l|c|c|c|c|}
\hline 
 & 16 & 32 & 64 & 128 \\
 & threads & threads & threads & threads \\
\hline %
{1 sub-domain}  & 10.6 & 16.7 & 25.4 & 20.2 \\
{2 sub-domains} & 11.9 & 22.1 & 34.3 & 45.4 \\
{4 sub-domains} & 10.4 & 22.3 & 35.1 & 50.7 \\
{8 sub-domains} & 11.2 & 24.2 & 39.8 & 66.9 \\
\hline 
\end{tabular}
}
\caption{Speedup of the Virtuelium DDM program (Ethernet) with respect to the number of threads and sub-domains.}
\label{tab:ddm_virtuelium}
\end{table}

\begin{table}
\centering
{\small
\begin{tabular}{|l|c|c|c|c|}
\hline 
 & 16 & 32 & 64 & 128 \\
 & threads & threads & threads & threads \\
\hline %
{1 sub-domain}  & 14.5 & 22.8 & 27.4 & 21.0 \\
{4 sub-domains} & 14.6 & 26.6 & 46.1 & 49.2 \\
{8 sub-domains} & 14.7 & 26.4 & 47.9 & 75.9 \\
\hline 
\end{tabular}
}
\caption{Speed-up of the acoustic DDM program (Ethernet) with respect to the number of threads and sub-domains.}
\label{tab:ddm_acoustic}
\end{table}

Speedups of Virtuelium execution are shown in table~\ref{tab:ddm_virtuelium}. They are very closed to what we obtained with the acoustic simulation software presented in table~\ref{tab:ddm_acoustic}~\cite{6636420}. Simulations were run on a hybrid, both distributed and shared memory, computational platform consisting of 4 nodes containing a quad core processor (a total of 16 cores). Each node were provided with 8 Gigabytes RAM (Random Access Memory). As we were expecting, DDM techniques significantly improved the performance of the parallelization. Although the speedups from acoustic simulation are quite better, we can notice that in both cases, from 16 to 128 threads, 8 sub-domains decomposition allowed to multiply the speedup by nearly 6, while classical parallelization only reach a factor less than 2. On another hand, for a fixed number of threads, the speedup keeps increasing as the number of sub-domains do.

\section{Conclusion}

In this paper, we proposed an original ray-tracing domain decomposition method for image rendering with natural lighting. According to domain decomposition methods principle, light rays characteristics have been matched as interface constraints between neighboring sub-domains. We presented a test case on a model of the church of the Royaumont abbey where we particularly deal with medieval glass material properties. It outlined the performance and efficiency of our method, relatively to multi-core architectures.

\section*{Acknowledgements}
The authors acknowledge the Foundation Royaumont for its help and in particular Jerome Johnson and Nathalie Le Gonidec for the helpful discussions and comments.

\bibliography{ref}
\bibliographystyle{abbrv}

\end{document}